# Molecular Orbital Electronic Instability in the van der Waals Kagomé Semiconductor Nb$_3$Cl$_8$: Exploring Future Directions

Yuya Haraguchi[1,2]* and Kazuyoshi Yoshimura[2,3]

[1]*Department of Applied Physics and Chemical Engineering, Tokyo University of Agriculture and Technology, Koganei, Tokyo 184-8588, Japan*
[2]*Department of Chemistry, Graduate School of Science, Kyoto University, Kyoto 606-8502, Japan*
[3]*Department of Energy and Hydrocarbon Chemistry, Graduate School of Engineering, Kyoto University, Nishikyo-ku, Kyoto 615-8510, Japan*

Nb$_3$Cl$_8$, a cluster Mott insulator with a distinctive magnetic molecular orbital structure organized into a breathing kagomé lattice, showcases critical phase transitions under specialized conditions. By transitioning from paramagnetic to nonmagnetic states below 90 K, we clarified this behavior through combined nuclear magnetic resonance and low-temperature X-ray diffraction studies, pointing to charge disproportionation as the driving force. Subsequent investigations via angle-resolved photoemission spectroscopy and first-principles calculations have disclosed topologically flat bands, confirming advanced electronic characteristics in Nb$_3$Cl$_8$. These discoveries not only deepen our comprehension of Mott insulators but also broaden our grasp of the dynamic interrelations among topology, electron interactions, and quantum phenomena in two-dimensional systems. The research on Nb$_3$Cl$_8$ thus lays foundational knowledge for advancing the exploration of quantum states in complex material systems, marking it as a critical model in the ongoing evolution of condensed matter physics.

## 1. Introduction

Recent research into two-dimensional (2D) materials has significantly enhanced our understanding of their potential in building future electronic, spintronic, and quantum technologies [1–7]. Van der Waals (vdW) compounds are particularly notable within this category, characterized by their weak interlayer and strong in-plane bonding, which contribute to their exceptional electronic and magnetic properties. These characters make them highly suitable for device integration. The seminal discovery of graphene highlighted the extraordinary electronic and mechanical properties inherent in 2D materials, provoking further research into this field. Moreover, studies on transition metal dichalcogenides (TMDs) have revealed exciting phenomena such as charge density waves and superconductivity, which further demonstrate the vast potential of vdW materials across various applications [8–16].

Recent advancements in the field of vdW magnets opened a new era in condensed matter physics, bringing forth discoveries that challenge longstanding theoretical concepts [4–7, 17–19]. Materials such as CrI$_3$ [20,21], Fe$_3$GeTe$_2$ [22–26], and Cr$_2$Ge$_2$Te$_6$ [27,28] have demonstrated ferromagnetic transitions in two-dimensional monolayers, contradicting the Mermin–Wagner theorem that suggests that continuous symmetries cannot break spontaneously at finite temperatures in low-dimensional systems. Additionally, breakthroughs such as the experimental realization of Ising antiferromagnetic transitions in FePS$_3$ monolayers, guided by Onsager's solution [29], underscore the ability of vdW magnets to redefine theoretical boundaries [30,31]. These materials also intersect with cutting-edge topics in quantum physics; MnBi$_2$Te$_4$ has been recognized as an antiferromagnetic topological insulator [32–34], and α-RuCl$_3$ has provided evidence of Majorana fermions [35–38], collectively enriching our understanding of the complex interplay between magnetic order and electronic states in two-dimensional systems.

These discoveries not only underscore the versatility and rich potential of vdW magnets as platforms for testing and expanding theoretical physics but also cement their status as central pillars in the ongoing exploration of novel quantum and magnetic phenomena in condensed matter physics. As research continues, vdW magnets are poised to offer more insights and potentially revolutionary findings in the field. Among these materials, Nb$_3$Cl$_8$ stands out owing to its unique electronic and magnetic transitions, which we will explore in detail.

Nb$_3$Cl$_8$, a vdW chloride with a layered structure, exemplifies the unique properties and challenges associated with these materials [39,40]. As a 2D vdW magnet, Nb$_3$Cl$_8$ not only continues the legacy of graphene and TMDs in enriching our understanding of low-dimensional systems but also introduces novel magnetic behaviors linked to its molecular orbital structure, further broadening the scope of research and applications in this dynamic field.

Nb$_3$Cl$_8$ is distinguished by not only its vdW structure but also its unique configuration of magnetic molecular orbitals, which assemble into a "breathing" kagomé lattice forming Nb$_3$ clusters. This configuration facilitates intriguing electronic interactions and phase behaviors. A key breakthrough in understanding Nb$_3$Cl$_8$ has been the discovery of charge disproportionation, a phenomenon where charges are transferred between molecular orbitals, thus stabilizing various magnetic and electronic states [41]. This has considerably enhanced our understanding of the material's low-temperature phase transitions from paramagnetic to nonmagnetic states, providing valuable insights into its underlying physical mechanisms.

Building upon these fundamental discoveries, recent advancements have revealed the presence of topological flat bands in $Nb_3Cl_8$, marking a significant development that parallels discoveries in metallic kagomé systems [42]. Despite being a semiconductor, $Nb_3Cl_8$ exhibits these flat bands as a result of its kagomé-like lattice structure, thereby introducing phenomena typically associated with metallic systems into a semiconducting context. This structural similarity to kagomé lattices has significantly heightened interest in $Nb_3Cl_8$, much like the attention garnered by kagomé metals noted for their exotic electronic properties [43–48].

The evolution of studies on $Nb_3Cl_8$, from the discovery of charge disproportionation to the exploration of its topological flat bands, marks significant progress in the study of two-dimensional materials. These advancements not only highlight the complex electronic structure of vdW magnets but also establish $Nb_3Cl_8$ as a crucial model for investigating quantum phenomena. This compound exemplifies the potential of vdW magnets to elucidate the effects of dimensionality on magnetism and electronic transport, particularly as its ability to be exfoliated to monolayer thickness allows for detailed studies on the effects of confinement and reduced dimensionality. Overall, $Nb_3Cl_8$ serves as a vital subject in the expanding field of advanced materials, offering insights into both fundamental properties and technological potentials.

In this review, we provide a comprehensive analysis of $Nb_3Cl_8$, highlighting its role as a prominent vdW magnets. We explore various aspects from synthesis techniques to structural properties, with a special emphasis on its paramagnetic to nonmagnetic phase transition analyzed using a charge disproportionation model. This model, along with our evaluation of different low-temperature crystallographic models, underpins our understanding of unique magnetic behavior in $Nb_3Cl_8$. Further advancements in the study of $Nb_3Cl_8$ include the discovery of topological flat bands, which have significantly broadened its application potential and cemented its status in quantum materials research. Additionally, the identification of $Nb_3Cl_8$ as a single-band Mott insulator opens promising avenues for researching quantum spin liquid states in its monolayer form [49,50], thus positioning $Nb_3Cl_8$ as a crucial material for future technological and scientific explorations.

By synthesizing past research and outlining prospective directions, we aim in this review to provide a comprehensive overview of the developments in the study of $Nb_3Cl_8$ and its significant role in advancing material science, particularly within the research field of quantum and 2D magnets.

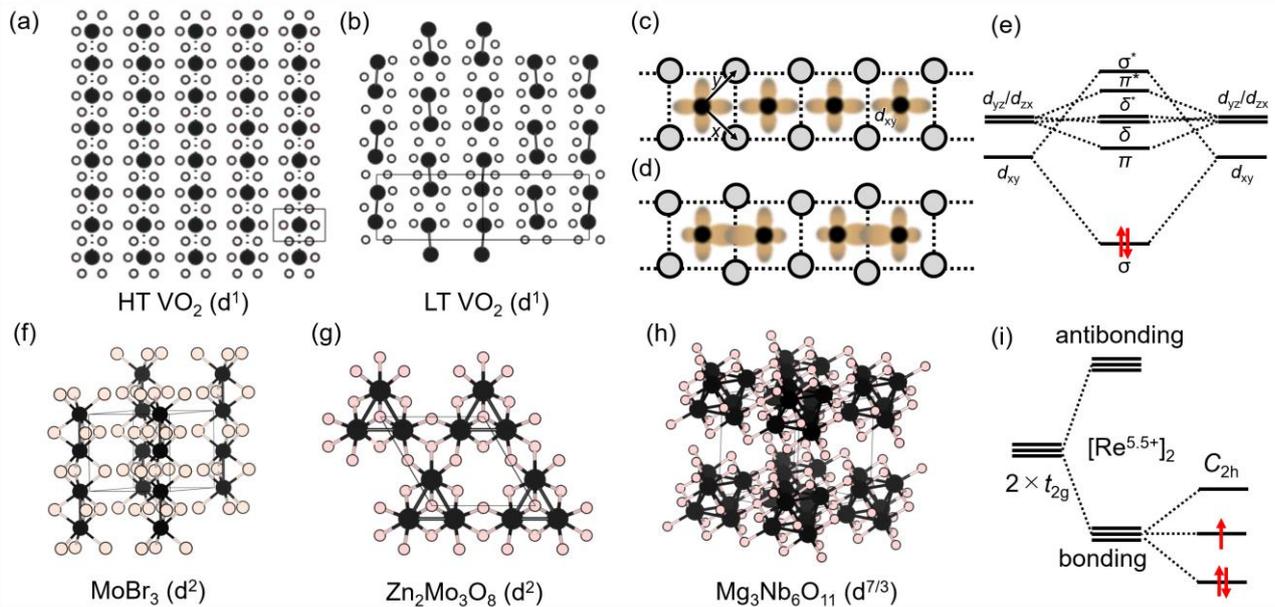

Fig. 1. (Color online) Crystallographic and molecular structural illustrations: (a) High-temperature rutile phase of $VO_2$, (b) low-temperature distorted phase of $VO_2$ with the formation of V–V dimers, (c) schematic of a regular chain consisting of edge-sharing octahedra, (d) schematic of a distorted chain with paired cations, (e) energy level diagram of molecular orbitals for a cation pair, (f) crystal structure of $MoBr_3$, (g) crystal structure of $Zn_2Mo_3O_8$, (h) crystal structure of $Mg_3Nb_6O_{11}$, and (i) electron configuration in molecular orbitals of half-integer mixed-valence dimers.

## 2. Electronic State of $Nb_3Cl_8$

*2.1 Cluster formation in common inorganic compounds*

In inorganic materials, the formation of cluster structures within compounds is a common phenomenon, often induced by lattice instabilities. These instabilities facilitate the stabilization of the system's energy through the formation of molecular orbitals among a selected group of magnetic ions. An illustrative example of this dynamic is found in a rutile type $VO_2$ [Fig. 1(a)].

The typical rutile structure features transition metal ions aligned in one-dimensional chains along the $c$-axis. In contrast, the low-temperature phase of $VO_2$ exhibits a distorted arrangement of $V^{4+}$ ions, resulting in two distinct types of V–V bond distances within these chains: 2.688 Å (short) and 3.304 Å (long) [Fig. 1(b)] [51–56]. This structural differentiation is pivotal in understanding its metal–insulator transition at 340 K. Such transitions in rutile-type oxides suggest a fundamental role for clustering in their phase transformations, as depicted in Figs. 1(c) and 1(d), which illustrate the dimerization of magnetic ions in a one-dimensional array [57].

The formation of short bonds between V atoms in $VO_2$ suggests the potential creation of metal–metal bonds, leading to molecular orbitals within the V–V dimer [58]. This process is schematically represented in Fig. 1(d), where $d^1$ magnetic ions form dimers, resulting in the loss of unpaired electrons and rendering the V–V dimer a nonmagnetic insulator. This mechanism of dimerization and subsequent electronic transformation extends beyond rutile compounds. For instance, $MoBr_3$ ($Mo^{3+}$: $d^3$) [59] also exhibits nonmagnetic insulating behavior owing to similar dimerization processes [Fig. 1(f)]. Moreover, compounds such as $Zn_2Mo_3O_8$ ($Mo^{4+}$: $d^2$) [60], which forms trimers with six d-electrons [Fig. 1(g)], and $Mg_3Nb_6O_{11}$ ($Nb^{8/3+}$: $d^{7/3}$) [61], which forms hexamers with fourteen d-electrons [Fig. 1(h)], further illustrate the diverse manifestations of clustering in inducing nonmagnetic insulating states.

Clustering within compounds is a prevalent phenomenon; however, the specific electronic configurations resulting from such structural formations can vary significantly depending on the valency requirements of the constituent ions. An illustrative example of this can be seen in compounds where transition metal cations with a formal mixed valence come together to form dimers, as depicted in Fig. 1(i). In this configuration, each dimer has odd d electrons to retain a single unpaired electron, leading to an $S = 1/2$ spin state for each cluster. Examples for $S = 1/2$ dimer include $Y_5Mo^{4.5+}_2O_{12}$ [62], $Ba_3LnIr^{4.5+}_2O_9$ ($Ln$ = lanthanoids) [63], and $La_3Re^{5.5+}_2O_{10}$ [64]. This distinctive electronic state implies that each cluster acts analogously to a magnetic ion, which opens versatile possibilities for designing magnets as "magnetic cluster material".

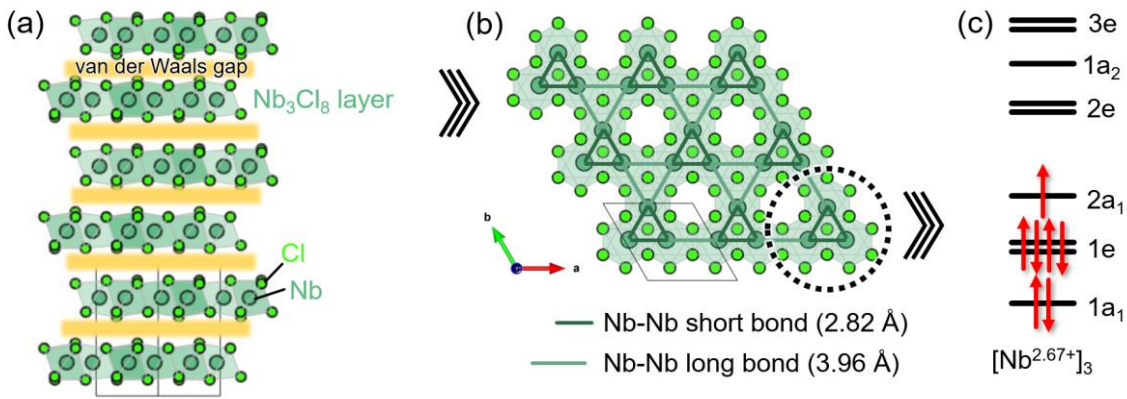

Fig. 2. (Color online) (a) Crystal structure of $Nb_3Cl_8$ including 2D layer with vdW gap. (b) $Nb_3Cl_8$ monolayer structure, where Nb-Nb bonds with different lengths are shown differently. Dark green lines indicate the Nb-Nb metallic bonds within each $Nb_3$ trimer. (c) Molecular orbital and electron configuration constructed in the $Nb_3$ trimer in $Nb_3Cl_8$.

*2.2 Crystal structure of $Nb_3Cl_8$*

$Nb_3Cl_8$ was identified a magnetic cluster material at the beginning of our research, prompting further investigation into its properties and behaviors. The structural analysis of $Nb_3Cl_8$ revealed that, as illustrated in Fig. 2(a), the compound is characterized by a vdW-type layered crystal structure [39,40]. The crystal structure of $Nb_3Cl_8$ bears similarities to that of $Zn_2Mo_3O_8$ [60], albeit with a significant modification: in $Nb_3Cl_8$, the $Mo_3O_8$ layered component of $Zn_2Mo_3O_8$ is substituted such that Mo sites are replaced by Nb and O sites by Cl. Additionally, the stacking pattern in $Nb_3Cl_8$ differs from that in $Zn_2Mo_3O_8$. Within each of these layers, the arrangement of niobium atoms forms what is described as a "breathing" kagomé lattice, as shown in Fig. 2(b). This unique structural feature includes two distinct types of metallic triangles, differentiated by their metal-metal bond distances, a notable deformation from the traditional, uniform kagomé lattice.

The "breathing" aspect of this lattice—referring to the alternating contraction and expansion of triangular units—plays a crucial role in the physical properties of the material [65–68]. The shorter triangular units within the lattice promote increased metal-metal bonding. This enhanced bonding is facilitated by the significant orbital overlap of d-electrons, which is a direct result of the closer proximity of the niobium atoms within these contracted triangles. Such orbital interactions are pivotal in defining the electronic and magnetic behaviors of the material, making the breathing modifications of the kagomé lattice a topic of substantial interest in the study of layered materials with potential exotic electronic states.

*2.3 Electronic state*

In the study of molecular orbitals in transition metal compounds, the case of $Nb_3Cl_8$ offers a compelling illustration of trimerization processes and their impact on electronic structure. This compound features niobium with a mixed valence of +2.67, resulting in an effective fictive valence of +8 per niobium trimer—a representation that, while unconventional, enhances the clarity of understanding in complex valence scenarios. In $Nb_3Cl_8$, the niobium trimers collectively share seven d-electrons, that populate the molecular orbitals as depicted in Fig. 2(c). The lowest energy states are filled first, leaving a single unpaired electron in the $2a_1$ orbital, thereby conferring a spin state of $S = 1/2$ on each niobium trimer. Given the arrangement of these trimers in a triangular lattice, $Nb_3Cl_8$ can be classified as an $S = 1/2$ triangular lattice magnetic system.

This molecular orbital framework finds parallels in molybdenum oxides such as $LiZn_2Mo_3O_8$ [69,70], $Li_2ScMo_3O_8$ [71,72], and $Na_3Sc_2(MoO_4)_2Mo_3O_8$ [73,74], which are rigorously investigated for their potential to exhibit quantum spin liquid (QSL) states because of their similar electron configurations and lattice structures [75–77].

Further examination of molecular orbitals in related systems reveals that molecules harboring unpaired electrons are typically less stable. For instance, in molybdenum oxide systems such as $M_2Mo_3O_8$ ($M$ = 3d transition metal) [78–81] and $LiAMo_3O_8$ (where $A$ is a rare earth ion) [82,83], molybdenum generally assumes a $Mo^{4+}$ oxidation state. This state results in each Mo trimer housing six d-electrons, thus leaving the $2a_1$ orbital vacant and making the $1e$ orbital the highest occupied molecular orbital (HOMO), which imparts a nonmagnetic character to these compounds.

An intriguing electronic modification is noted in $Nb_3TeCl_7$, as outlined in Ref. 84, where replacing one-eighth of the $Cl^-$ ions in $Nb_3Cl_8$ with $Te^{2-}$ ions shifts the oxidation state of niobium from $Nb^{2.67+}$ to $Nb^{3+}$. This oxidation change reduces the d-electron count in the molecular orbitals to six per Nb trimer, effectively emptying the $2a_1$ orbital and rendering the compound nonmagnetic. Conversely, in $Nb_3Cl_8$, the intercalation of sodium leads to the formation of $NaNb_3Cl_8$, which lowers the oxidation state of niobium from $Nb^{2.67+}$ to $Nb^{2.33+}$. This decrease results in eight d-electrons per Nb trimer, fully occupying the $2a_1$ orbital, thus maintaining a nonmagnetic state. These modifications underscore the pivotal role of electron pairing in molecular orbitals, which not only stabilizes the material but also resolves valence instabilities, consistently leading to nonmagnetic outcomes.

The observations related to $Nb_3Cl_8$ reveal that this compound hosts one unpaired electron in its molecular orbital, positioning it as a "magnetic molecular orbital crystal." This structure is characterized by a triangular lattice of crystallized molecular orbitals. Such a configuration is particularly noteworthy because it suggests a framework where exotic electronic states could emerge, primarily due to the charge instability associated with the unpaired electron. The unique arrangement of molecular orbitals in a triangular lattice not only enhances the magnetic instability but also introduces the potential for charge fluctuation phenomena, which are pivotal in the development of novel electronic states.

## 3. Crystal Growth and Preparation of Polycrystalline Sample

In the synthesis of $Nb_3Cl_8$, initial attempts involved mixing elemental niobium and niobium pentachloride ($NbCl_5$) in a stoichiometric Nb:Cl ratio of 3:8 in an argon-filled glove box. The mixture was subsequently pressed into pellets and sealed in a vacuum within a quartz tube, then subjected to a heat treatment at 700 °C for 48 h. This method, however, resulted in a material that exhibited mechanical fragility, disintegrating easily upon handling, thereby proving unsuitable for producing high-quality polycrystalline samples via conventional solid-state reaction methods.

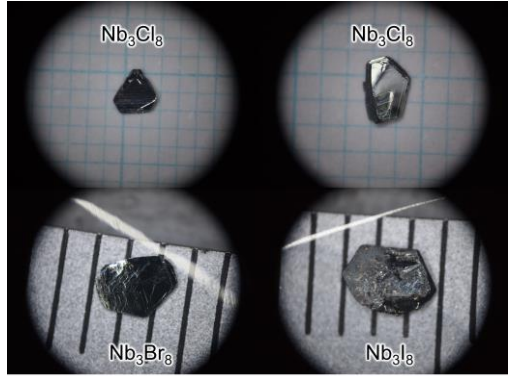

Fig. 3. (Color online) Photographs of single-crystal $Nb_3Cl_8$, $Nb_3Br_8$, and $Nb_3I_8$. $Nb_3Cl_8$ crystals are displayed on 1 mm×1 mm grid paper to emphasize scale, while $Nb_3Br_8$ and $Nb_3I_8$ are shown against a ruler with 1 mm increments.

Given the challenges encountered with the initial material, it was repurposed as a precursor for single-crystal growth. Traditional vapor transport techniques were initially employed but failed to yield crystals of adequate size for magnetic property measurements. A more successful approach involved the use of a flux growth method, which operated under a horizontal configuration that maintained a temperature gradient along the axis. This technique, bearing resemblance to liquid transport growth, involved the use of lead (II) chloride ($PbCl_2$) as the flux medium [41]. The precursor powder mixed with $PbCl_2$ was sealed under argon atmosphere in a horizontally placed quartz tube within a tubular furnace. The temperature settings were optimized with the hot end at 750 °C and the cold end at 700 °C over 150 hours, followed by a controlled cooling rate of about 10 °C per hour. After the growth process, the residual $PbCl_2$ adhering to the single-crystal specimens was effectively removed by soaking in hot water. Subsequent research has expanded upon initial methodologies, with recent experiments reporting the successful growth of single crystals using the chemical vapor phase transport method. Notably, these advancements utilized ammonium chloride ($NH_4Cl$) as the transport agent [85].

The resultant high-quality polycrystalline samples of $Nb_3Cl_8$ were then obtained by ultrasonically treating these single crystals. Representative single crystals, illustrated at the top of Fig. 3, were used for the property measurements later reported in our previous work [41]. By a similar growth methodology, single crystals of $Nb_3Br_8$ were also successfully cultivated. Interestingly, single crystals of $Nb_3I_8$ were obtained serendipitously during the initial attempts to synthesize polycrystalline material, without the need for transport agents.

The crystalline structure of these materials, particularly $Nb_3Cl_8$, exhibits pronounced cleavability along the ab plane, allowing for easy exfoliation with simple tools such as Scotch tape—similar to the exfoliation of graphite. This property suggests that the layers within $Nb_3Cl_8$ are interconnected primarily through weak vdW forces, a characteristic that might contribute to their unique electronic and magnetic properties.

## 4. Basic Physical Properties

*4.1 Magnetism*

Figure 4(a) shows a comparative analysis of the temperature-dependent magnetic susceptibility χ for single-crystal $Nb_3Cl_8$, alongside $Nb_3Br_8$ and $Nb_3I_8$, offering insights into their magnetic behaviors. In the high-temperature regime, both $Nb_3Cl_8$ and $Nb_3Br_8$ exhibit behavior consistent with the $S = 1/2$ Curie–Weiss law, corroborating the electronic configurations illustrated in Fig. 2(c). The Weiss temperatures are $\theta_w = -13.1$ K for $Nb_3Cl_8$ and $-6.1$ K for $Nb_3Br_8$, indicating weak antiferromagnetic coupling among the unpaired electrons in the cluster orbitals. Notably, $Nb_3Cl_8$ undergoes a pronounced first-order phase transition accompanied by significant hysteresis at around 90 K, and $Nb_3Br_8$ at around 380 K, with the transition temperatures and magnitudes remaining consistent across different magnetic field orientations.

This leads to a hypothesis that $Nb_3Br_8$, which is nonmagnetic at room temperature and crystallizes in the $R\bar{3}m$ space group—a configuration different from $Nb_3Cl_8$ at the same temperature—undergoes a similar structural phase transition from $P\bar{3}1m$ to $R3$. Additionally, systematic studies in the $Nb_3Cl_{8-x}Br_x$ solid-solution series have demonstrated a continuous increase in this phase transition temperature with an increase in Br concentration [86], highlighting the effect of compositional tuning on structural and magnetic properties.

Further exploration is presented in Fig. 4(b), which details the magnetic susceptibility of powdered $Nb_3Cl_8$ samples derived via the ultrasonic treatment of single crystals. Unlike their single-crystal counterparts, these powdered samples did not display the primary phase transition, maintaining a paramagnetic state down to approximately 2 K [41]. This absence suggests that the primary phase transition in $Nb_3Cl_8$ is not intrinsic to individual layers but rather stems from interlayer coupling, with the size effect induced by ultrasonic exfoliation affecting this interlayer interaction.

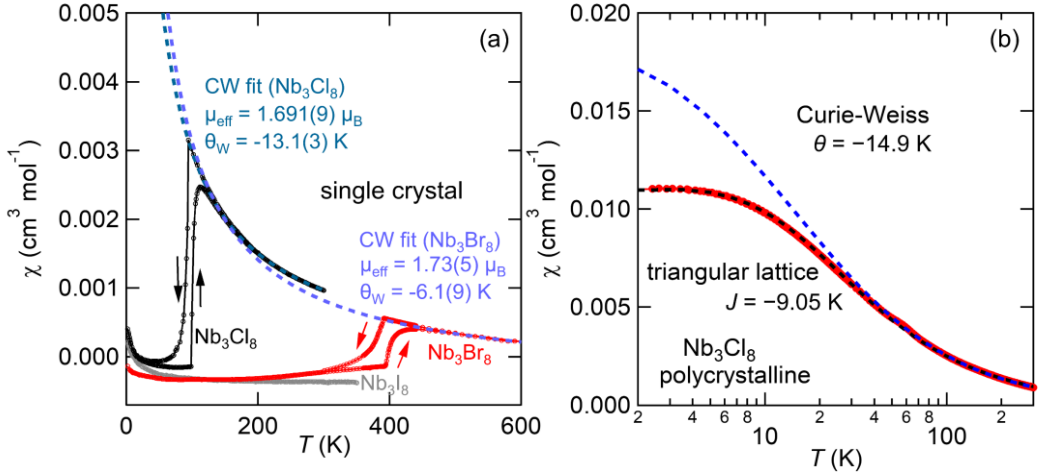

Fig. 4. (Color online) (a) Temperature-dependent magnetic susceptibility of $Nb_3Cl_8$ single crystals, with comparative data for $Nb_3Br_8$ and $Nb_3I_8$. The dotted line represents a Curie–Weiss fit, while up and down arrows mark the direction of the temperature sweep. (b) Magnetic susceptibility of a polycrystalline $Nb_3Cl_8$ sample prepared via ultrasonic treatment, shown with both a Curie-Weiss fit and a fit to the $S = 1/2$ triangular lattice Heisenberg model.

Moreover, the deviations from the Curie-Weiss law observed in these powdered samples at low temperatures align with predictions of the two-dimensional triangular lattice antiferromagnetic Heisenberg model, with the coupling constant $J = -9.1$ K. This confirms the realization of the two-dimensional triangular lattice electronic state proposed in Fig. 2. Ongoing discussions around the potential emergence of a spin liquid state in $Nb_3Cl_8$ further underscore the material's suitability for exploring two-dimensional frustrated magnetism [41].

While $Nb_3I_8$ might be expected to undergo a phase transition at temperatures exceeding those observed for $Nb_3Br_8$, investigations into this compound have been halted owing to concerns about its thermal stability. Specifically, at elevated temperatures, there is a pronounced risk that $Nb_3I_8$ could decompose, possibly leading to the release of iodine gas. The potential for such decomposition not only raises safety concerns but also poses a significant risk of damaging experimental equipment.

*4.2 Structural phase transition*

In an effort to deepen our understanding of the structural nuances of the nonmagnetic state in single-crystal $Nb_3Cl_8$, we have employed X-ray diffraction techniques on crystals at both 300 and 24 K; the structural outcomes are depicted in Fig. 5. The analysis at 300 K confirms the congruence with structures previously documented, affirming the consistency and reliability of the experimental setup and crystal quality [41].

A notable aspect of this study is the observed symmetry reduction from $P\bar{3}1m$ to $R3$, coinciding with a temperature-driven phase transition. This transition results in a tripling of the periodicity along the c-axis in the low-temperature phase relative to the high-temperature counterpart. This alteration is further accompanied by a shift in the relative positioning of the $Nb_3$ clusters, with layer shifts approximately in the $[-1/3, 1/3, 0]$ direction, as evidenced at high temperatures (Fig. 5).

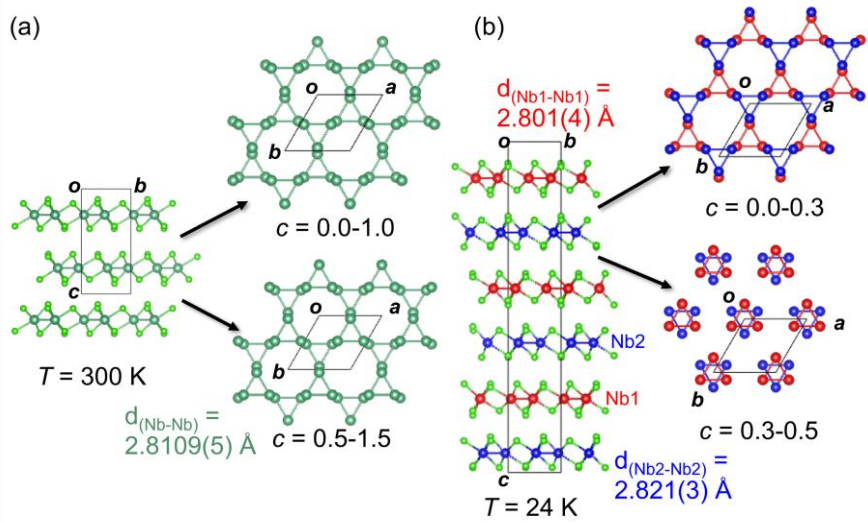

Fig. 5.   (Color online) Comparative crystal structures of $Nb_3Cl_8$ at 300 K (a) and 24 K (b), illustrating the structural changes that occur with temperature, highlighting the Nb–Nb bond lengths within the $Nb_3$ trimers at each temperature.

One intriguing finding is the potential suppression of this structural phase transition in polycrystalline samples, possibly due to size effects, suggesting that the physical properties of $Nb_3Cl_8$ could be finely tuned through careful control of the crystal structure.

Additionally, in the low-temperature phase, each unit cell maintains an equivalent number of Nb sites, a fact that aligns well with nuclear magnetic resonance (NMR) spectroscopy results. The phase transition modifies the Nb-Nb bond distances within the trimers, transitioning from a uniform $d$(Nb-Nb) = 2.8109 Å at 300 K [Fig. 5(a)] to differentiated bond lengths of $d$(Nb1-Nb1) = 2.801 Å and $d$(Nb2-Nb2) = 2.821 Å at 24 K [Fig. 5(b)]. This structural variation manifests in the stacking of the crystal, where layers of $Nb(1)_3Cl_8$ and $Nb(2)_3Cl_8$ stack alternately, each characterized by distinct intercluster bond distances. This layer alternation and the associated modulation of bonding distances underscore the complex interplay between structural and electronic properties in $Nb_3Cl_8$, revealing potential avenues for further exploration of phase-dependent phenomena in such materials [41].

*4.3 Charge disproportionation model*

The charge disproportionation model, as detailed in Fig. 6, offers a robust explanation for the observed changes in cluster bond lengths and the demagnetization occurring during the phase transition at around $T^* \sim 90$ K in $Nb_3Cl_8$. Initially, in the high-temperature phase, the $Nb_3$ clusters maintain uniform bond lengths and exhibit a collective valence of $Nb^{2.67+}$, corresponding to seven d electrons per cluster. This configuration results in a paramagnetic state characterized by an unpaired electron in the $2a_1$ orbital, indicative of an unstable electronic system.

As the temperature decreases, a charge disproportionation process is initiated to stabilize the electronic system. This transition, from $d^7 + d^7$ molecular orbitals to a combination of $d^6 + d^8$ molecular orbitals, facilitates the pairing of unpaired electrons in the $2a_1$ orbital of one cluster, while leaving the $2a_1$ orbital of the other cluster empty, effectively realizing a nonmagnetic state.

A key aspect of this discussion is the variation in Nb−Cl bond lengths within the $NbCl_6$ octahedra. For the $Nb_3$ trimer, the Nb–Cl bond length is expected to increase from $[Nb^{2.67+}]_3$ to $[Nb^{2.33+}]_3$ and decrease from $[Nb^{2.67+}]_3$ to $[Nb^{3+}]_3$. The average length of the Nb–Cl bond $\langle d_{Nb-Cl} \rangle$ in the $NbCl_6$ octahedron is expected to increase with the valence state of niobium. Following charge disproportionation in the $Nb_3$ cluster, $\langle d_{Nb1-Cl} \rangle$ decreases from 2.499 to 2.492 Å, while $\langle d_{Nb2-Cl} \rangle$ increases from 2.499 to 2.507 Å. These changes in bond lengths align with the charge disproportionation model.

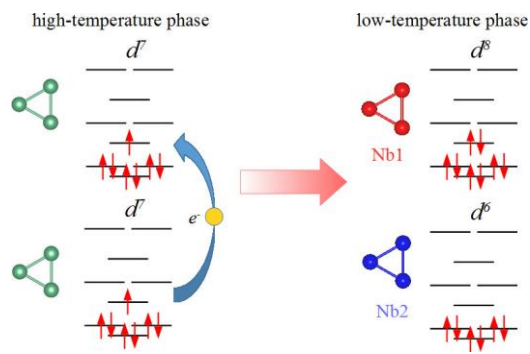

Fig. 6.  (Color online) Schematic view of the charge disproportionation of the Nb$_3$ trimers at $T^* = 90$ K in Nb$_3$Cl$_8$. Reprinted (adapted) with permission from Ref. 41. Copyright {2017} American Chemical Society.

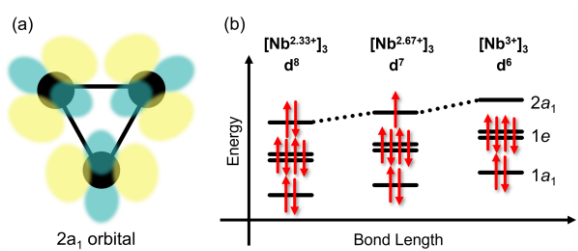

Fig. 7.  (Color online) (a) Schematic view of the $2a_1$ molecular orbital of the Nb$_3$ trimer. (b) Schematic view of energy levels in the bonding orbitals of Nb$_3$ trimers with corresponding bond lengths and with their electron configurations.

The impact of these bond length modifications extends to the energy levels of the $2a_1$ orbital, the primary bonding orbital in this context. The expansion or contraction of the cluster bonding length directly affects the energy level of this orbital, as depicted schematically in the molecular orbital view of the Nb$_3$ trimer shown in Fig. 7(a). In the [Nb$^{2.67+}$]$_3$ clusters with a $d^7$ configuration, the $2a_1$ orbital exhibits an occupation-energy gain. Specifically, in the [Nb$^{2.33+}$]$_3$ clusters with a $d^8$ configuration, the energy-lowering $2a_1$ orbital becomes fully occupied. In contrast, in the [Nb$^{3+}$]$_3$ clusters with a $d^6$ configuration, the energy-lowering $2a_1$ orbital remains unoccupied, as shown in Fig. 7(b). Consequently, the observed variations in cluster bonding length provide compelling evidence for charge disproportionation induced by interlayer charge transfer [41].

## 5.  Different Crystal Structural Models

In a parallel study to ours, the McQueen group conducted structural analyses of Nb$_3$Cl$_8$ using low-temperature X-ray diffraction (XRD) on powder samples [87]. Their findings highlighted a phase transition under nonmagnetic conditions, identifying a transition to a $C2/m$ crystal structure. Their report offers a valuable comparison for examining the structural dynamics of Nb$_3$Cl$_8$.

We assess both the similarities and differences between the $R3$ and $C2/m$ structures reported in the literature. Such comparative analysis is instrumental in understanding the structural complexities and phase behavior of Nb$_3$Cl$_8$ under various conditions. In addition, our own investigations employing low-temperature XRD and magnetization measurements on powder samples of Nb$_3$Cl$_8$ revealed no evidence of a structural phase transition, which is consistent with the magnetization measurement using sonicated powdered Nb$_3$Cl$_8$.

In the study of Nb$_3$Cl$_8$'s structural phases, the high-temperature structure, characterized by the space group $P\bar{3}m1$, consists of 22 atoms per unit cell, as illustrated in Fig. 8(a). Upon undergoing a structural phase transition, this configuration evolves into more complex low-temperature structures. Specifically, the $C2/m$ and $R3$ structures, observed at lower temperatures, contain 44 and 66 atoms per unit cell, respectively. This increase in the number of atoms is primarily attributed to changes in the orientation of the unit cell, which is a direct consequence of the phase transition.

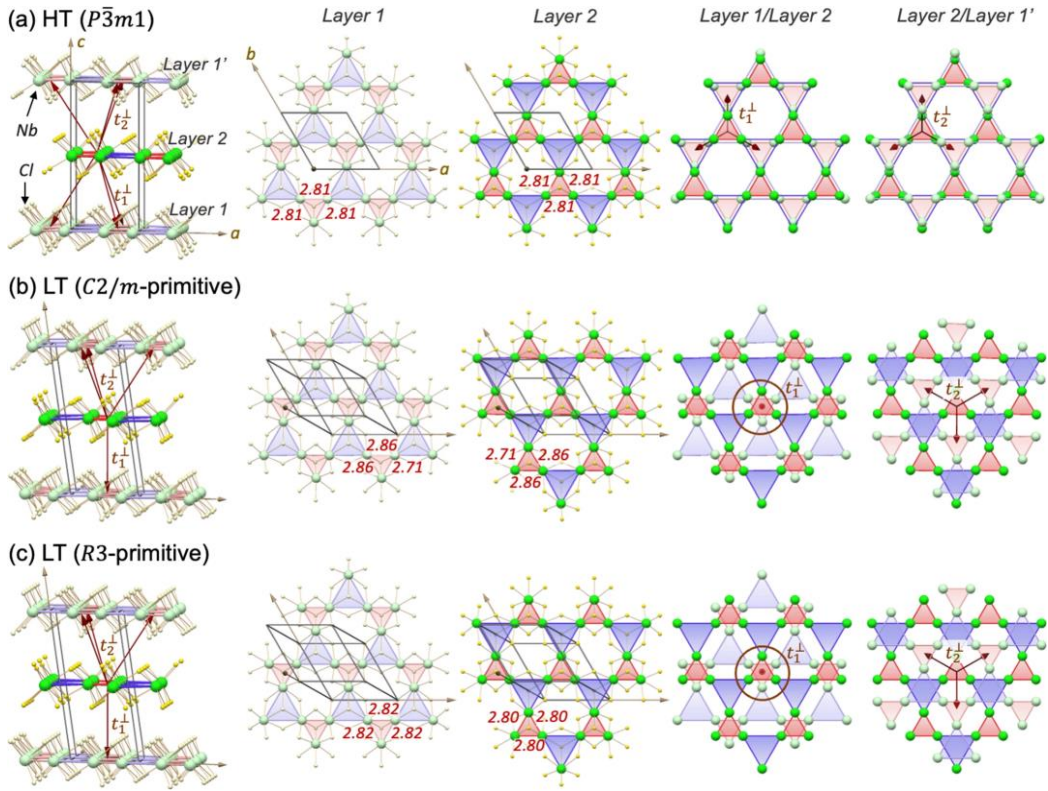

Fig. 8. (Color online) Crystal structures of bulk $Nb_3Cl_8$: (a) at high temperature (space group $P\bar{3}m1$) and (b, c) at low temperatures (space groups $C2/m$ and $R3$, respectively). Top views of alternating layers are shown with atoms depicted in faded and bright colors (green and yellow) to distinguish Nb and Cl atoms in adjacent layers. Brown arrows highlight interlayer couplings between nearest $Nb_3$ trimers. Distances between nearest Nb atoms in each layer are marked in angstroms (Å). Reprinted (adapted) with permission from Ref. 49. Copyright {2024} Springer Nature.

To facilitate a clearer comparison between these structures, a primitive cell representation containing 22 atoms, akin to the high-temperature phase, is employed in Figs. 8(b) and 8(c). Within this simplified model, the $C2/m$ and $R3$ structures are shown to be nearly equivalent. The transition from one structure to the other is driven by a shift in the $Nb_3Cl_8$ monolayer from its original high-temperature arrangement. The fundamental difference between the $C2/m$ and $R3$ configurations lies in the presence or absence of threefold symmetry. This distinction is crucial for understanding the symmetry-related changes that define the structural phase transitions in $Nb_3Cl_8$.

In the comparative structural analysis of $Nb_3Cl_8$ polymorphs, one notable aspect is the variation in the allocation of niobium sites across different structures. The $C2/m$ structure is characterized by distinct site occupations with Nb1 located at the $4i$ site and Nb2 at the $8j$ site, establishing an atomic ratio of 1:2. In contrast, the $R3$ structure provides a more balanced configuration, with both Nb1 and Nb2 occupying the $9b$ sites, resulting in an equimolar ratio of 1:1. This variance in site distribution plays a pivotal role in influencing the physical properties of these phases.

Furthermore, $Nb_3Br_8$, known to be nonmagnetic at room temperature, is typically indexed in the $R\bar{3}m$ space group [88]. The challenge in differentiating $R3$ and $R\bar{3}m$ structures in X-ray diffraction (XRD) studies arises owing to the equivalence of their Bragg reflection indices. However, detailed single-crystal XRD analyses of $Nb_3Cl_8$ suggest that while $R\bar{3}m$ indexing is possible, the $R3$ model offers a more accurate description. This model, which indicates differing Nb-Nb bond distances at the two niobium sites, correlates well with NMR spectroscopy findings [41]. The presence of two distinct niobium sites in the $R3$ structure contrasts sharply with the singular site model of $R\bar{3}m$, making $R3$ a more suitable structural interpretation given the NMR data. In the structural characterization of $Nb_3Cl_8$ within the $R3$ model, a noteworthy feature emerges concerning the arrangement of niobium sites across the layers. This model reveals that different Nb sites are alternately positioned within each layer, a pattern that is vividly illustrated in the model diagrams with colors red and blue, signifying the spatial inversion asymmetry and the breaking of mirror symmetry inherent in this structure. Despite these localized differences, it is important to note that the overarching stacking pattern of the $Nb_3Cl_8$ monolayers maintains a uniform consistency throughout the crystal (Fig. 9).

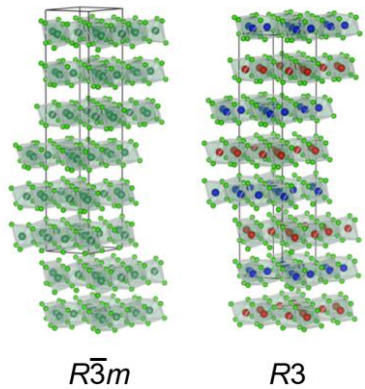

Fig. 9. Comparison between $R\bar{3}m$ and $R3$ structural models of Nb$_3$Cl$_8$. The $R3$ model distinctly exhibits two different niobium sites, represented in red and blue, which alternate within each layer, thereby breaking mirror symmetry and showcasing spatial inversion asymmetry contrasting with the $R\bar{3}m$ model, which assumes a singular niobium site.

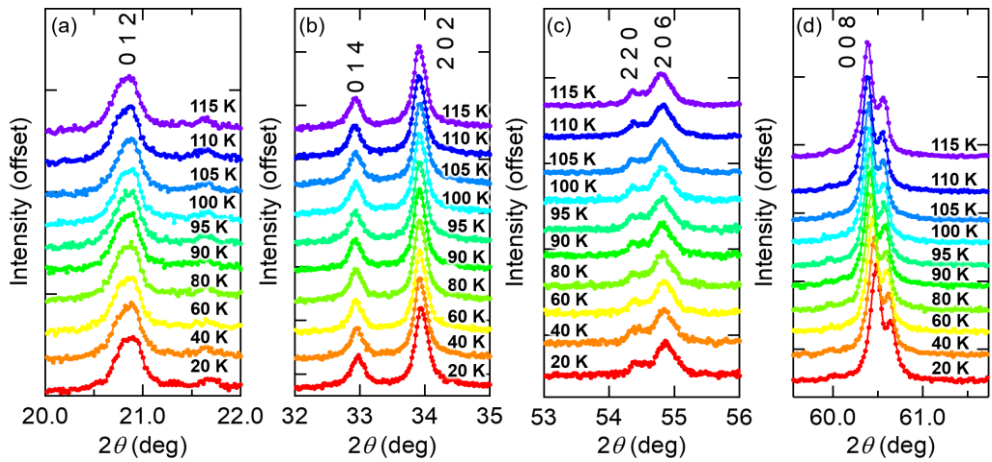

Fig. 10. (Color online) Temperature evolution of the powder X-ray diffraction (XRD) profiles for Nb$_3$Cl$_8$, obtained using Cu-K$\alpha$1 radiation on samples prepared by sonicating powdered crystals. The diffraction data includes several indexed Bragg reflections corresponding to the high-temperature $P\bar{3}m1$ structure. For enhanced clarity, each diffraction curve is systematically shifted downward by specific offsets.

The determination of the room-temperature structure of Nb$_3$Br$_8$ remains elusive, primarily owing to the challenges in obtaining single crystals robust enough for comprehensive XRD analysis. This ongoing uncertainty has spurred efforts to apply alternative techniques, such as NMR and second harmonic generation (SHG) experiments, to ascertain whether the room-temperature structure aligns with the previously reported $R\bar{3}m$ or if it exhibits the spatial inversion symmetry breaking characteristic of the $R3$ structure.

In addition, recent Raman spectroscopy studies have provided pivotal insights, particularly those affirming the persistence of threefold symmetry in its low-temperature phase, which supports the $R3$ crystal structure model rather than the monoclinic $C2/m$ model [85]. Polarized Raman spectroscopy and magneto-Raman techniques were employed to meticulously analyze the phonon modes of Nb$_3$Cl$_8$ as it changes from high to low temperatures. Notably, these experiments demonstrated that the phonon mode symmetries remain consistent through the temperature change, indicating the retention of the $R3$ symmetry. This finding decisively challenges earlier assertions favoring a shift to the $C2/m$ symmetry, which would involve a deviation from the inherent threefold rotational symmetry of the $R3$ structure.

To assess the validity of the proposed differences in the structural model of Nb$_3$Cl$_8$, we implemented low-temperature powder X-ray diffraction (XRD) measurements on a sonicated powder sample, which was previously observed not to undergo a structural phase transition at around the expected temperature of $T^* \sim 90$ K, as detailed in Fig. 10. Notably, the analysis did not reveal any peak splitting above or below $T^*$, thereby providing substantial evidence that no such transition occurred within the powdered $Nb_3Cl_8$ sample.

Moreover, the absence of any indicators of a paramagnetic to nonmagnetic phase transition corroborates these findings. This consistency underscores the potential limitations of using low-temperature XRD on powder samples for detecting subtle structural transitions, especially in the absence of magnetic phase changes.

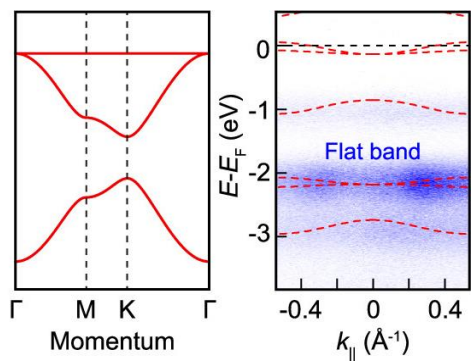

Fig. 11.  (Color online) Identification of topological flat bands (TFBs) and a moderate band gap in $Nb_3Cl_8$ through a combination of first-principles calculations and angle-resolved photoemission spectroscopy (ARPES) measurements. Reprinted (adapted) with permission from Ref. 42. Copyright {2022} American Chemical Society.

## 6.  Development and Future Prospects

*6.1 A platform for exploring kagomé topological flat bands*

Recent research in our collaboration work investigated by Sun *et al.* has unveiled the significant role of topological flat bands (TFBs) within the kagomé lattice of $Nb_3Cl_8$, shedding light on the novel physical properties these structures can induce [42]. Subsequent studies by other research groups have also documented the observation of TFBs in $Nb_3Cl_8$ [89].

The study focuses on $Nb_3Cl_8$, a kagomé semiconductor that distinguishes itself by combining these topological bands with a semiconducting behavior, creating a moderate band gap that is crucial for device applications. Using advanced techniques such as angle-resolved photoemission spectroscopy (ARPES) and first-principles calculations, researchers have observed the presence of TFBs alongside a clear band gap of approximately 1.12 eV, as shown in Fig. 11. This discovery is pivotal as it addresses the challenge of achieving an 'OFF' state in kagomé-lattice-based devices, which is typically hindered in metallic kagomé materials [42].

Furthermore, the ability to mechanically exfoliate $Nb_3Cl_8$ into monolayers that remain stable under ambient conditions adds to its applicability in developing quantum and optoelectronic devices [42]. This stability, combined with the semiconducting properties and the presence of a magnetic ground state in monolayer $Nb_3Cl_8$, opens up avenues for exploring interactions between geometry, topology, and magnetism in two-dimensional materials.

The identification and characterization of TFBs in $Nb_3Cl_8$ not only enhance our understanding of topological insulators but also introduce potential applications in fabricating devices that leverage the unique properties of kagomé lattices [90–92]. This could lead to the development of new technologies that utilize the quantum mechanical effects inherent in these topological structures. In addition, the experimental results reveal that $Nb_3Cl_8$ retains a flat band profile despite undergoing trimerization, which is a process where three adjacent Nb atoms cluster, significantly affecting the electronic structure [93]. This trimerization forms an effective triangular lattice, laying the groundwork for the hypothesized QSL state. Intriguingly, the low-temperature magnetic susceptibility data show transitions that hint at a QSL phase, characterized by fluctuating spins that evade crystallization into a rigid order, thereby preventing long-range magnetic order down to very low temperatures.

This line of inquiry has further catalyzed the discovery of TFBs in related compounds such as $Nb_3TeCl_7$ [94], $Nb_3I_8$ [95,96], and $Nb_3Br_8$ [97], expanding the scope and excitement within this research field. These findings not only enrich our understanding of the unique electronic properties of these materials but also highlight the broader applicability and significance of topological phenomena in similar layered compounds. Furthermore, computational chemistry is playing a pivotal role in expanding the exploration of novel materials. For instance, substances such as $W_3Cl_8$ [98] and $Co_3X_8$ ($X$ = Cl, Br) [99], which are predicted to have structures similar to that of $Nb_3Cl_8$, have been identified through computational models, broadening the horizons for material research. The emergence of TFBs across this series suggests a rich tapestry of electronic behaviors awaiting exploration, underpinning the growing interest and ongoing developments in the study of topological materials.

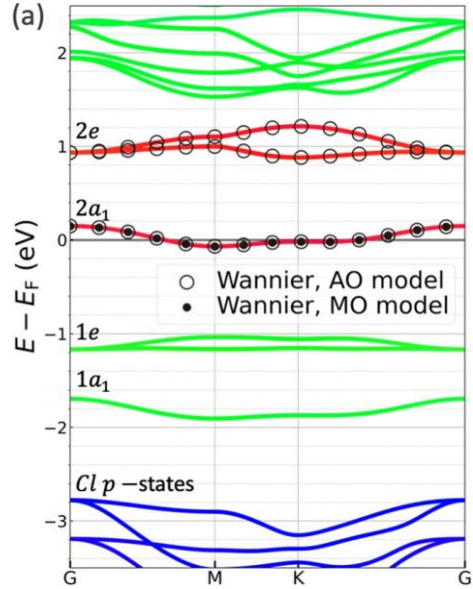

Fig. 12. DFT band structure of $Nb_3Cl_8$. Reprinted (adapted) with permission from Ref. 49. Copyright {2024} Springer Nature.

*6.2 Ideal realization of the Hubbard model in monolayer $Nb_3Cl_8$*

As reported by Grytsiuk *et al.*, monolayer $Nb_3Cl_8$ has been identified as a prototypical Mott–Hubbard insulator, which exemplifies the Hubbard model in a real material [49]. Initially conceptualized for simple theoretical analyses of electronic correlations in solids, the Hubbard model traditionally simplifies electron behaviors into single-electron hopping and on-site Coulomb repulsion within a lattice. Despite its simplicity, applying this model to real materials has proved challenging owing to the complex nature of electronic interactions and the influence of multiple orbitals at the Fermi level in most materials.

In $Nb_3Cl_8$, a unique electronic structure featuring a well-defined single band crossing the Fermi level, as determined through density functional theory (DFT) calculations and experimental observations, provides an ideal basis for applying a single-orbital Hubbard model. This model simplifies the complex electronic interactions within $Nb_3Cl_8$, focusing on its Mott insulating state characterized by a significant energy gap, which is dependent on environmental conditions.

Through advanced computational methods such as the constrained random phase approximation (cRPA) [100] reveal how the isolated electronic band and the localized Coulomb interactions support a robust Mott insulator phase in the monolayer $Nb_3Cl_8$ as shown in Fig. 12. This is substantiated by comparing theoretical predictions with experimental data, including ARPES [42] and transport measurements [101], which confirm the presence of a gap indicative of strong electronic correlations.

Additionally, in this study, it is discussed that the modifications in electronic properties brought about by changes in the dielectric environment, highlighting the sensitivity of the Mott insulating state to external influences. This aspect is critical for understanding the material's behavior in practical applications, where environmental factors can significantly affect performance [102].

The results of this research underscore the significance of $Nb_3Cl_8$ as a nearly perfect realization of the Hubbard model in a two-dimensional material, offering a platform for exploring fundamental electronic phenomena and the potential for innovative electronic devices based on Mott insulators. Further investigations into the low-temperature phases and bulk properties of $Nb_3Cl_8$ promise to expand our understanding of strongly correlated materials and the practical applications of Hubbard model predictions in material science and technology.

*6.3 Possible quantum spin liquid*

Our investigations have revealed that the powdered $Nb_3Cl_8$ suppresses the structural phase transition observed in its crystalline form, as shown in Fig. 2(b). This suppression significantly affects the magnetic properties of the powder sample, as evidenced by our experimental observations of magnetic susceptibility, aligning well with theoretical predictions based on the triangular lattice Heisenberg model. This suggests a stabilization of the high-temperature phase in powdered forms, which is a finding that diverges from behaviors observed in more ordered sample preparations.

Recent studies on $Nb_3Cl_8$ by Liu *et al.* highlight its potential to host a QSL state owing to its unique two-dimensional triangular lattice of Nb$_3$ trimers [50]. The cluster configuration leads to a net magnetic moment and introduces significant frustration, potentially stabilizing a QSL state. Notably, Nb$_3$Cl$_8$ undergoes a transition to a nonmagnetic state below approximately 90 K owing to a structural change, preventing direct observation of its magnetic ground state in single crystals.

Experiments conducted on this material, including specific heat measurements and magnetic susceptibility analysis, indicate unusual low-temperature properties consistent with QSL expectations. The specific heat displays a linear temperature dependence at low temperatures, suggestive of the presence of spinon Fermi surfaces—a hallmark of QSL states. This characteristic is further supported by the temperature-independent behavior of the magnetic susceptibility at low temperatures. The persistence of the high-temperature magnetic phase in ground powders or samples compressed along the *c*-axis challenges the conventional understanding of phase transitions in such systems. These findings propose that the magnetic properties of Nb$_3$Cl$_8$, particularly under conditions that inhibit its structural transition, may provide fertile ground for exploring QSL states.

Exploring the potential for a QSL in a monolayer of Nb$_3$Cl$_8$ presents a compelling avenue of research, particularly when considered in relation to similar behaviors observed in compounds such as LiZn$_2$Mo$_3$O$_8$ [69,70], Li$_2$ScMo$_3$O$_8$ [71,72], and Na$_3$Sc$_2$(MoO$_4$)$_2$Mo$_3$O$_8$ [73,74]. These materials are known for their QSL behavior [75-77], which are characterized by cluster-based electronic states similar to those proposed for Nb$_3$Cl$_8$. The comparison of these materials provides an excellent opportunity to deepen our understanding of the conditions under which QSLs can emerge and persist.

The exploration of QSL states in Nb$_3$Cl$_8$ and its comparison with other molybdates that exhibit analogous electronic properties could shed light on the intricate interplay between lattice geometry, electronic interactions, and magnetic frustration. Such studies are crucial for elucidating the mechanisms that stabilize QSLs and could pave the way for new theoretical and experimental frameworks in the study of low-dimensional quantum phenomena.

*6.4 Pressure-induced symmetry breaking*

Recent studies on Nb$_3$Cl$_8$ under high pressure have revealed significant structural and electronic transformations [103]. Initially characterized by a trigonal structure with a homogeneously distributed electron cloud within Nb$_3$ clusters, Nb$_3$Cl$_8$ undergoes a dramatic phase transition when subjected to pressures nearing 8 GPa. This transition involves a shift from the trigonal $P\bar{3}m1$ to a monoclinic $C2/m$ symmetry, resulting in the breaking of threefold rotational symmetry and the emergence of electron disproportionation within the Nb$_3$ clusters.

This structural shift is accompanied by a redistribution of electrons from a delocalized state (Nb$^{8/3+}$ for each atom within the cluster) to a localized state, characterized distinctively by Nb$^{3+}$, Nb$^{3+}$, and Nb$^{2+}$ configurations. Such electron disproportionation fundamentally alters the electronic properties of the material, as evidenced by changes in electrical transport and UV-Vis absorption measurements [103]. These alterations result in a subtle bandgap discontinuity, signaling a shift in the electronic structure without a transition to metallic behavior, a rarity under compression scenarios which typically favor metallization.

In addition, we consider the potential similarities between the high-pressure $C2/m$ structure and the McQueen group's structural model [87]. In our investigations, the mechanical crushing of Nb$_3$Cl$_8$ single crystals in a mortar is accompanied by a faint smell of chlorine, hinting at slight decomposition of the sample. To mitigate this, ultrasonic treatment is employed as an alternative method for pulverizing the single crystal Nb$_3$Cl$_8$ samples. This approach not only prevents potential decomposition but also raises an intriguing possibility: different preparation methods might stabilize distinct structural forms in powder samples. This observation underscores the importance of methodological choices in the study of structural properties under varying conditions.

## 7. Summary

In this review, we explore the significant contributions that studies of Nb$_3$Cl$_8$ have made to the field of vdW materials and two-dimensional magnetic systems. Our examination reveals profound insights into its distinctive magnetic molecular orbital crystal structure, which forms a breathing kagomé lattice exhibiting remarkable electronic and magnetic properties.

A focal point of our discussion is the paramagnetic to nonmagnetic transition observed in Nb$_3$Cl$_8$ at low temperatures, a process driven by charge disproportionation. This phenomenon, involving the redistribution of charges among the molecular orbitals of the Nb$_3$ trimer clusters, critically affects the material's magnetic and electronic properties, marking a significant deviation from typical behaviors observed in similar systems.

Furthermore, the discovery of TFBs in Nb$_3$Cl$_8$ represents a pivotal advancement, greatly expanding the potential applications of this material and solidifying its importance in the field of quantum materials research. These flat bands arise from the specific geometry of its triangular lattice, reminiscent of kagomé structures, which positions Nb$_3$Cl$_8$ as a candidate for exhibiting exotic quantum states traditionally associated with metallic systems, despite its semiconducting nature.

Moreover, experimental variations in conditions such as temperature and pressure have yielded deep insights into the effect of these factors on the stability and expression of different magnetic and electronic phases in Nb$_3$Cl$_8$. Understanding these dynamics is crucial for effectively linking theoretical models to empirical data, thereby enhancing our understanding of how intrinsic properties such as lattice structure and electronic interactions can govern the behavior of two-dimensional materials.

In this review, we synthesized these critical developments, positioning $Nb_3Cl_8$ as an essential material for ongoing research in condensed matter physics, especially in the study of vdW and two-dimensional magnets. The insights gained from $Nb_3Cl_8$ not only enrich our understanding of its unique properties but also contribute to broader discussions on novel materials capable of hosting QSL states and other complex phenomena. As research on $Nb_3Cl_8$ continues, its diverse properties promise to open new avenues for technological innovation and deepen the fundamental understanding of materials science.

## Acknowledgment

This work was supported by JST PRESTO Grant No. JPMJPR23Q8 and JSPS KAKENHI Grant Nos. JP23H04616 (Grant-in-Aid for Transformative Research Areas (A) "Supra-ceramics"), JP22K14002, JP19K14646 (Grant-in-Aid for Early-Career Scientists), and JP16J04048 (Grant-in-Aid for JSPS Fellows).

*E-mail: chiyuya3@go.tuat.ac.jp